\documentclass[12pt]{IEEEtran}
\onecolumn
\usepackage{setspace}
\usepackage{lineno,hyperref}

\usepackage{multirow}
\usepackage{hhline}
\usepackage{soul,color}
\usepackage{amsmath}
\usepackage{suffix,mathtools}

\usepackage{suffix,mathtools}
\DeclarePairedDelimiterX\MeijerM[3]{\lparen}{\rparen}%
{#3\,\delimsize\vert\begin{smallmatrix}#1 \\ #2\end{smallmatrix}}

\newcommand\MeijerG[8][]{%
  G^{\,#2,#3}_{#4,#5}\MeijerM[#1]{#6}{#7}{#8}}

\WithSuffix\newcommand\MeijerG*[7]{%
  G^{\,#1,#2}_{#3,#4}\MeijerM*{#5}{#6}{#7}}

\begin{document}


\title{Error Rate and Ergodic Capacity of RF-FSO System with Partial Relay Selection in the Presence of Pointing Errors}

\author{Milica Petkovic,~
        Imran~Shafique~Ansari,
        Goran~T.~Djordjevic, Khalid~A.~Qaraqe
 \thanks{M.~Petkovic is with University of Novi Sad, Faculty of Technical Science, 21000 Novi Sad, Serbia (e-mail:  milica.petkovic@uns.ac.rs).}
 \thanks{I. S. Ansari is with  School of Engineering, University of Glasgow, Glasgow G12 8QQ, United Kingdom.}
 \thanks{G. T. Djordjevic is with University of Nis, Faculty of Electronic Engineering, Department of Telecommunications, A. Medvedeva 14, 18000 Nis, Serbia.}
 \thanks{K. A. Qaraqe is with Electrical and Computer Engineering (ECEN) Department, Texas A\&M University at Qatar (TAMUQ), Education City, Doha, Qatar.}
}

\maketitle

\begin{abstract}
This paper presents an analysis of a multiple dual-hop relaying system, which is composed of km-class radio frequency (RF)-free-space optical (FSO) links. Partial relay selection based on outdated channel state information (CSI) is employed in order to select active relay for further transmission. Amplify-and-forward relaying protocol is utilized. The RF links are assumed to be subject to Rayleigh fading, and the FSO links are influenced by both Gamma-Gamma atmospheric turbulence and pointing errors. On the basis of our  previously derived expression for cumulative distribution function of the equivalent signal-to-noise ratio of the whole system, we derive novel  analytical expressions for the average bit-error rate (BER) and ergodic capacity that are presented in terms of  the Meijer's \textit{G}-function and extended generalized bivariate Meijer's \textit{G}-function, respectively. The numerical results are confirmed by Monte Carlo simulations. Considering the effect of time-correlation between outdated CSI and actual CSI related to the RF channel at the time of transmission, the average BER and the ergodic capacity dependence on various system and channel parameters are observed and discussed. The results illustrate that the temporal correlation between outdated and actual CSI has
strong effect on system performance, particularly on BER values,
when FSO hop is influenced by favorable conditions.
\end{abstract}




\section{Introduction}
\label{SecI}

Radio frequency (RF) systems, which are very often used for backhaul networking, cannot support high data rates of great number of users and other requirements of the $5^{ \rm th}$ generation wireless networks  \cite{R1n}. Because of that, free-space (FSO) optical  systems have been adopted as a complement or alternative  to the radio frequency (RF) technology,  especially in overcoming the connectivity hole between the main backbone system and last mile access network. The use of FSO systems provides a license-free and high data rates transmission \cite{magazin1, magazin2,R3,R4,R5,R6,R7}. The FSO links are valuable for enabling a large number of RF users to be multiplexed through a single FSO link. 
Ciaramella \textit{et al.} \cite{G} realized an FSO system between two buildings in Pisa, Italy, and reached the date rate of 1.28 Tb/s over a distance of 210 m. The recent experiment, carried out by German Aerospace Center (DLR), proved that data rate of 1.72 Tb/s can be achieved over an FSO link with length of 10.45 km  \cite{S}. These experimental demonstrations proved that FSO could be a promising technology for achieving high quality-of-services and high data rates in 5G networks.

The main reason for the intensity fluctuations of the received optical signal is  atmospheric turbulence, which occurs as a result of the variations in atmospheric altitude, temperature, and pressure. The misalignment between the transmitter laser and the detector at the receiver (called pointing errors) is another cause of intensity fluctuations of the optical signal. Although received optical signal fluctuations can be mitigated by diversity techniques \cite{bookMATLAB}, and a number of techniques have been developed to ensure alignment between transmitter and receiver of FSO link \cite{novaknjiga}, these two effects have attracted attention of many researchers \cite{PE,PE1,PE2,PE_MISO}. The Gamma-Gamma distribution is widely adopted  for modeling intensity fluctuations due to atmospheric turbulence \cite{bookMATLAB,Andrews,Habash,Survey,GG}, while the pointing errors are described by the model derived with an assumption that the  total  radial displacement at the receiver detector is subject to Rayleigh distribution \cite{PE,PE1,PE2,PE_MISO}. That means transmitter and receiver are aligned initially perfectly, but due to beam wandering and building sway, tracking is not perfect and random misalignment appears. This assumption is relevant for FSO links with km-class lengths.

The main challenge in the FSO link implementation is the obligatory  presence of the line-of-sight (LOS) between FSO apertures. Since the realization of this LOS requirement is quite challenging in some scenarios (difficult terrains such as crowded urban streets and areas), the idea of utilizing relaying technology  within FSO systems has been arised to accomplish coverage area extension. More precisely, the  mixed (asymmetric) dual-hop amplify-and-forward (AF) RF-FSO relaying system, composed of RF and FSO links, was firstly introduced in \cite{Lee}. In order to perform electrical-to-optical signal conversion at the relay, subcarrier intensity modulation (SIM) technique can be applied \cite{bookMATLAB,end-end}. In addition, the mixed RF-FSO systems  enable multiple RF users to be multiplexed via a single FSO link \cite{Ansari-Impact}.
Besides \cite{Lee}, the performance analysis  of the asymmetric  RF-FSO systems with employing fixed AF gain relay was investigated in \cite{end-end, Ansari-Impact, Ansari-VTCSpring, Ansari-SIECPC, Anees1, Anees2, Zhang_JLT, Zedini_WCNC, fixnew1, Zedini_PhotonJ}.  Contrary, the performance of the asymmetric RF-FSO systems with employing variable AF gain relay was presented in \cite{Zedini_PhotonJ, JSAC1, JSAC2, Ansari-VTCFall, IWOW2013}, while \cite{DF} considered decode-and-forward RF-FSO system. Additionally, the impact of the interference at relay on the overall system performance was investigated in \cite{int1,int2}. The multiuser RF-FSO system was analyzed  in \cite{mu1,mu2,mu3,mu4}. In order to expand range and improve the performance limitations of FSO communications,  triple-hop RF/FSO/RF communication system was proposed in \cite{tri1,tri2}.

With aim to improve the system performance, implementation of multiple relays in RF  systems were widely investigated in past literature \cite{RF1, RF1a, RF2, RF3, RF4, RF5, RF6, RF7}. In order to avoid additional network delays and to achieve power savings, the partial relay selection (PRS) was   introduced in \cite{RF2}, considering the scenario when the active relay is chosen on the basis of single-hop instantaneous channel state information (CSI). 

The idea of PRS procedure utilization in the asymmetric  RF-FSO systems  employing fixed AF relays was proposed in \cite{JLT}, wherein the first RF hops experience Rayleigh fading, and the second FSO hops are affected by  the Gamma-Gamma atmospheric turbulence. In addition, the impact of the pointing errors on the same system was observed in \cite{chapter}, providing the  novel expressions for the outage probability. In \cite{dGG}, the multiple relayed mixed RF-FSO system with PRS was analyzed, but the FSO link was influenced by Double Generalized Gamma atmospheric turbulence. Furthermore,  performance analysis of the RF-FSO system with multiple relays was performed in \cite{HI1,HI2}, taking into account hardware impairments.

The previous studies \cite{chapter, dGG} was concentrated on determining outage probability. However, besides outage probability, other performance metrics are also important.  From both users' and designers' point of view, it is very important to know the probability that a bit transmitted over a channel will be wrongly detected, known as bit error rate.  In addition, we are focused on determining the maximum  data rate that could be supported by a channel when error probability can be downscaled under arbitrary low value. In this work, we extend the analysis from \cite{chapter} to estimating of ergodic capacity and average bit error rate (BER). Although the system model is quite similar compared with the one presented in \cite{chapter}, analytical derivations are completely  novel, and numerical results have not been previously reported. Novel analytical expressions for the average BER and the ergodic capacity are derived in terms of the Meijer's \textit{G}-function and the extended generalized bivariate Meijer's \textit{G}-function (EGBMGF), respectively. These expressions are utilized for examining some interesting effects of FSO and RF channels parameters  on overall system performance.  The analysis is carried out in the case when RF intermediate-frequency signal is amplified and modulated in optical carrier.

The remainder of the paper is organized in the following way. Channel and system models are described in Section~\ref{SecII}.  Section~\ref{SecIII} gives the average BER and the ergodic capacity analysis. Numerical results and simulations with corresponding comments are given in Section~\ref{SecIV}, while Section~\ref{SecV} concludes the paper.

\section{System and channel model}
\label{SecII}

\begin{figure}[!b]
\centering
\includegraphics[width=4in]{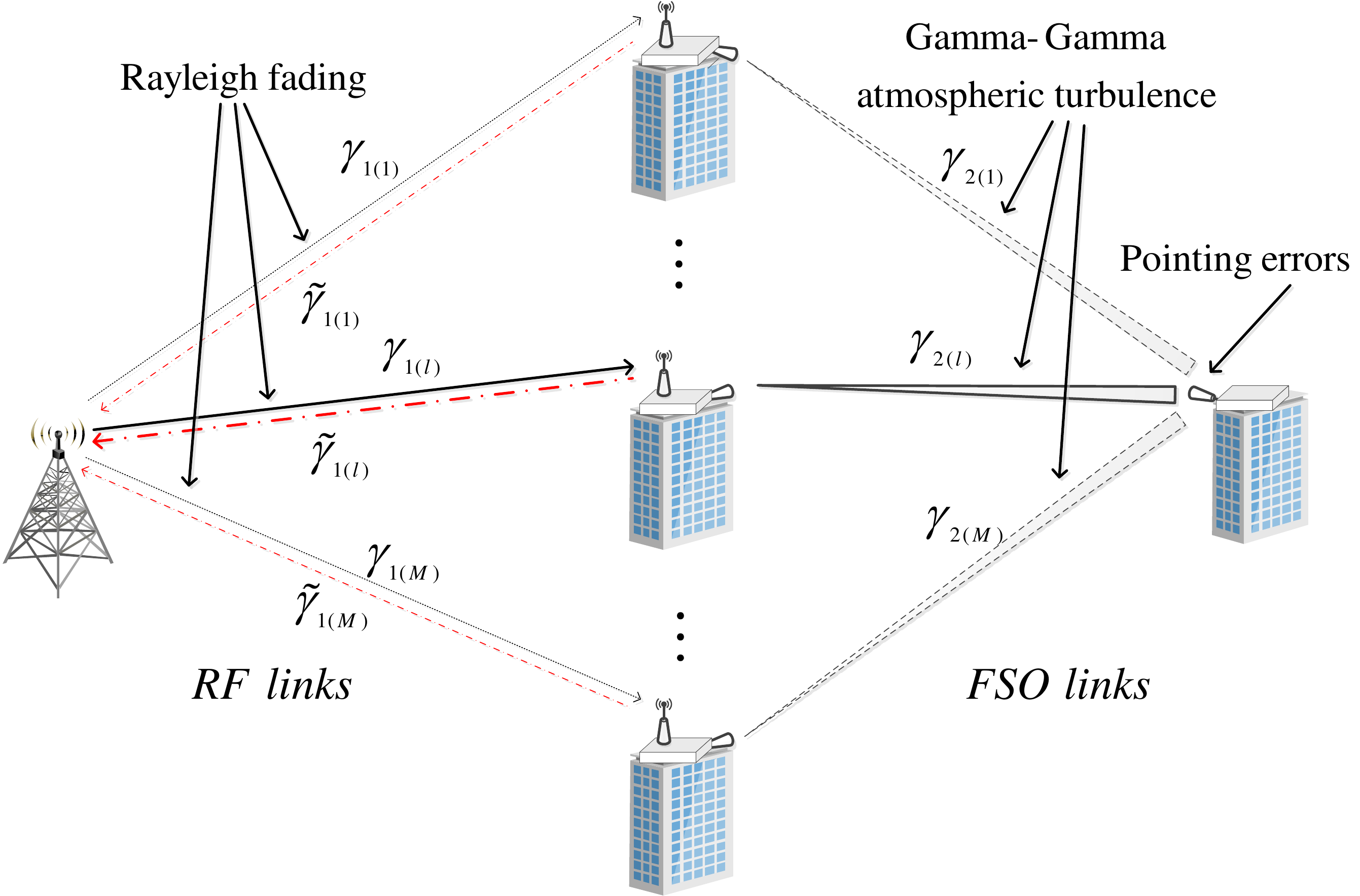}
\caption{Mixed RF-FSO system with PRS}
\label{Fig_1}
\end{figure}

The paper presents the analysis of the RF-FSO relaying system, assuming that the signal transmission from source to the active relay is performed in frequency range from 900 MHz to 2.4 GHz.  Asymmetric AF dual-hop RF-FSO system, presented in Fig.~\ref{Fig_1}, consists of  source $ S $, destination $ D $, and $ M \geq 1 $ relays, assuming there is no direct link between $ S $ and $ D $ nodes. Based on the local feedback sent from  the relays, the source node  $ S $ monitors the conditions of the first RF hops, and selects the active relay for further transmission via FSO channel. Active relay is selected as the best one
on the basis of estimated CSIs of the RF hops. Since time-varying nature of the RF hops is usual in practical scenarios, and due to feedback delay, the estimated CSI is not the same as actual one at the time of signal transmission. Because of that reason, following analysis considers the estimated CSI as outdated and time-correlated with the actual CSI of the RF hop. In addition, the selected active relay  is not maybe  available. In that case, the source chooses the next best relay, etc., and the  PRS procedure is performed via the $ l $th worst (or  $(M -l)$th best) relay $ R_{(l)} $ \cite{RF5}.

After the active relay selection, signal is transmitted over the selected RF hop. The electrical signal at the $ l $th relay is defined as
\begin{equation}
{r_{R(l)}} = {h_{SR(l)}}r + {n_{SR}},
\label{signal_r}
\end{equation}
where $ r $ represents a complex-valued baseband representation of the RF signal sent from the source node $ S $ with an average power $ P_s $, $ h_{SR(l)} $ is the fading amplitude over the $ S-R_{(l)} $ hop with $ {\rm E}\left[ {h_{SR(l)}^2} \right] = 1 $,   $ ({\rm E}\left[  \cdot  \right] $ denotes mathematical expectation$ ) $,  and $ n_{SR} $ denotes an additive white Gaussian noise with zero mean and variance  $ \sigma _{SR}^2$.

Based on  (\ref{signal_r}), the instantaneous signal-to-noise ratio (SNR) of the first RF hop is defined as
\begin{equation}
{\gamma _{1(l)}} = \frac{{h_{SR(l)}^2{P_s}}}{{\sigma _{SR}^2}} = h_{SR(l)}^2{\mu _1},
\label{SNR_1}
\end{equation}
where $ \mu_1 $ is the average SNR defined as  $ {\mu _1}~=~{{{P_s}} \mathord{\left/
 {\vphantom {{{P_s}} {\sigma _{SR}^2}}} \right.
 \kern-\nulldelimiterspace} {\sigma _{SR}^2}}. $
  
The signal  $ r_{R(l)} $ is amplified by the fixed gain $ G $ at the relay. The amplification is performed based on long-term statistic of the first RF hop. In this case, the relay gain $ G $ is determined as \cite{RF6}
\begin{equation}
{G^2} = \frac{1}{{{\rm E}\left[ {h_{SR(l)}^2} \right]\!\!{P_s} + \sigma _{SR}^2}}=\frac{1}{{\sigma _{SR}^2\left( {{\rm E}\left[ {{\gamma _{1(l)}}} \right] + 1} \right)}}\!=\!\frac{1}{{\sigma _{SR}^2\Re}},
\label{gain}
\end{equation}
where $ \Re= {\rm E}\left[ {{\gamma _{1(l)}}} \right] + 1$.

The amplified signal modulates an optical source (laser) intensity. The non-negativity requirement is ensured by adding dc bias. The optical signal at the relay output is given by
\begin{equation}
{r_{opt}} = {P_t}\left( {1 + Gm{r_{R(l)}}} \right),
\label{r_opt}
\end{equation}
where $ P_t $ denotes transmitted optical power and $ m $ is the modulation index $ (m=1) $. Signal is transmitted via free space and collected by the receiving telescope. Direct detection is performed and dc bias is removed. PIN photodetector is employed to perform an optical-to-electrical signal conversion. The electrical signal at the node $ D $ is given by
\begin{equation}
\begin{split}
{r_{D(l)}}& = {I_{R(l)D}}\eta {P_t}G{r_{R(l)}} + {n_{RD}} \\
& \!=\! {I_{R(l)D}}\eta {P_t}G{h_{SR(l)}}r + {I_{R(l)D}}\eta {P_t}G{n_{SR}} + {n_{RD}}, 
\end{split}  
\label{r_D}
\end{equation}
where $ {I_{R(l)D}} $ represents the received optical signal intensity, and $ n_{RD} $ represents the thermal noise modeled by the Gaussian distribution with zero mean and variance  $ \sigma _{RD}^2 $. An optical-to-electrical conversion coefficient is denoted by $ \eta$.

Based on (\ref{gain}) and (\ref{r_D}), the overall SNR at the destination is \cite{chapter}
\begin{equation}
{\gamma _{eq}} = \frac{{I_{R(l)D}^2{\eta ^2}P_t^2{G^2}h_{SR(l)}^2{P_s}}}{{I_{R(l)D}^2{\eta ^2}P_t^2{G^2}\sigma _{SR}^2 + \sigma _{RD}^2}} = \frac{{{\gamma _{1(l)}}{\gamma _{2(l)}}}}{{{\gamma _{2(l)}} + \Re}},
\label{SNR_eq}
\end{equation}
where $ \gamma _{2(l)} $ represents the instantaneous SNR over FSO link, given by
\begin{equation}
{\gamma _{2(l)}} = \frac{{I_{R(l)D}^2{\eta ^2}P_t^2}}{{\sigma _{RD}^2}}.
\label{SNR_2}
\end{equation}
The electrical SNR over FSO link is defined as  $ \mu _2={{{\eta ^2}P_t^2{{\rm E}^2}\left[ {{I_{R(l)D}}} \right]} \mathord{\left/
 {\vphantom {{{\eta ^2}P_t^2{{\rm E}^2}\left[ {{I_{R(l)D}}} \right]} {\sigma _{RD}^2}}} \right.
 \kern-\nulldelimiterspace} {\sigma _{RD}^2}} $.

\subsection{RF channel model}
The source monitors the conditions of the first RF hops by local feedbacks sent from relays. The active relay is selected based on the estimated CSIs of all RF hops. The estimated CSIs are assumed to be outdated and time-correlated with the actual corresponding CSIs of the RF hops.
Furthermore, the fact that the best selected relay is not necessarily available for further transmission is taken into consideration.

The RF hops are subject to Rayleigh fading. The probability density function (PDF) of the instantaneous SNR per RF hop between the source  and the $ l $th relay is derived in detail in \cite{RF5, JLT, chapter}, and is given by
\begin{equation}
\begin{split}
 {f_{{\gamma _{1(l)}}}}\left( x \right) = l{M \choose l}\sum\limits_{i = 0}^{l - 1} {l-1 \choose i}  \frac{(-1)^i}{{\mu _1((M - l + i)(1 - \rho ) + 1)}}{e^{ - \frac{{(M - l + i + 1)x}}{{\left( {(M - l + i)(1 - \rho ) + 1} \right){\mu _1}}}}}, 
\end{split} 
\label{PDF_SNR_1}
\end{equation}
where $\rho$ represents correlation coefficient between the instantaneous SNR over RF hop at the time of transmission ($ \gamma _{1(l)} $) and its outdated estimated version ($ \tilde \gamma  _{1(l)} $), which is used for relay selection. 

The cumulative distribution function (CDF) of $ \gamma _{1(l)} $  is 
\begin{equation}
\begin{split}
{F_{{\gamma _{1(l)}}}}(x)&=1 - l{M \choose l}\sum\limits_{i = 0}^{l - 1} {l-1 \choose i}  
\frac{{{{( - 1)}^i}}}{{(M - l + i + 1)}}{e^{ - \frac{{(M - l + i + 1)x}}{{\left( {(M - l + i)(1 - \rho ) + 1} \right){\mu _1}}}}}. 
\end{split} 
\label{CDF_SNR_1}
\end{equation}
The constant $ \Re $ is found by (\ref{gain}) and (\ref{PDF_SNR_1}) as \cite[(6)]{RF6}
\begin{equation}
\begin{split}
 \Re= 1 + l{M \choose l}\sum\limits_{i = 0}^{l - 1} {l-1 \choose i} \frac{{{{( - 1)}^i}\left( {(M - l + i)(1 - \rho ) + 1} \right){\mu _1}}}{{{{(M - l + i + 1)}^2}}}. \hfill \\ 
\end{split}
\label{const}
\end{equation}

\subsection{FSO channel model}
The considered system assumes that the intensity fluctuations of optical signal at the destination originate from the Gamma-Gamma atmospheric turbulence and pointing errors. The PDF of $ I_{R(l)D} $ is  \cite[(12)]{PE1}
\begin{equation}
\begin{split}
{f_{{I_{R(l)D}}}}({I_{R(l)D}})= \frac{{{\psi ^2}\alpha \beta }}{{{A_0}\Gamma (\alpha )\Gamma (\beta )}} \MeijerG*{3}{0}{1}{3}{\psi^2}{\psi^2 - 1, \, \alpha - 1, \, \beta - 1}{\dfrac{\alpha \beta}{A_0 } I_{R(l)D}}, 
\end{split} 
\label{PDF_I}
\end{equation}
where $ G_{p,q}^{m,n}\left(  \cdot  \right) $ is Meijer's \textit{G}-function \cite[(9.301)]{Grad}. 
Note that received signal variations due to scintillation and pointing errors are taken into account by (\ref{PDF_I}). The deterministic path loss due to scattering and diffraction \cite{bookMATLAB, novaknjiga} can be straightforwardly included. The path loss is relevant in the case when results should be presented in terms of the radiated optical power.
The  parameters $ \alpha $ and $ \beta $ are used to define an effective numbers of the scattering environment small-scale and large-scale cells, respectively, which are, with the assumption of the plane wave propagation and zero inner scale, defined as
\begin{equation}
\begin{split}
  \alpha  = {\left( {\exp \left[ {{{0.49\chi_R^2} \mathord{\left/
 {\vphantom {{0.49\chi_R^2} {{{\left( {1 + 1.11\chi_R^{12/5}} \right)}^{7/6}}}}} \right.
 \kern-\nulldelimiterspace} {{{\left( {1 + 1.11\chi_R^{12/5}} \right)}^{7/6}}}}} \right] - 1} \right)^{ - 1}}, \hfill \\
  \beta  = {\left( {\exp \left[ {{{0.51\chi_R^2} \mathord{\left/
 {\vphantom {{0.51\chi _R^2} {{{\left( {1 + 0.69\chi _R^{12/5}} \right)}^{5/6}}}}} \right.
 \kern-\nulldelimiterspace} {{{\left( {1 + 0.69\chi _R^{12/5}} \right)}^{5/6}}}}} \right] - 1} \right)^{ - 1}}. \hfill \\ 
\end{split} 
\label{alfa i beta}
\end{equation}
The Rytov variance is defined as  $ \chi_R^{2}=~1.23C_n^{2}\iota^{7/6}d^{11/6} $,   $ \iota = 2\pi/\lambda $ represents the wave number with the wavelength $ \lambda $, and $ d $ is the   FSO link length. The refractive index structure parameter is denoted by $ C_n^{2} $, varying in the range from $ 10^{-17} $ to $ 10^{-13} $ m$ ^{-2/3} $ for weak to strong turbulence.
The  parameter relating to the pointing errors, $ \psi $, is defined as
\begin{equation}
\psi  = \frac{{{a_{{d_{eq}}}}}}{{2{\sigma _s}}},
 \label{ksi}
\end{equation}
where  $ {a_{{d_{eq}}}} $ is the equivalent beam radius at the receiver and $ \sigma_s $ represents the pointing error (jitter) standard deviation at the receiver. The parameter $ {a_{{d_{eq}}}} $  is dependent on the beam radius at the distance $ d $, $ a_d $, as $a_{{d_{eq}}}^2=~a_d^2\sqrt \pi {{\operatorname{erf} (v)} \mathord{\left/
 {\vphantom {{erf(v)} {(2v\exp ( - {v^2}))}}} \right.
 \kern-\nulldelimiterspace} {(2v\exp ( - {v^2}))}} $ , with $ v =~{{\sqrt \pi  a} \mathord{\left/
 {\vphantom {{\sqrt \pi  a} {(\sqrt 2 {a_d})}}} \right.
 \kern-\nulldelimiterspace} {(\sqrt 2 {a_d})}} $ and the parameter $ a$ being the radius of a circular detector aperture. 
The parameter $ A_0 $ is defined as  $ {A_0} = {\left[ {\operatorname{erf} \left( v \right)} \right]^2} $, where $ \operatorname{erf} \left(  \cdot  \right) $ is the error function \cite[(8.250.1)]{Grad}.

\newcounter{MYtempeqncnt}
\begin{table*}[!t]

\normalsize
\setcounter{MYtempeqncnt}{\value{equation}}
\centering
\label{tab:label}
\caption{Constants and system and channels parameters (unless otherwise is stated)}

\begin{tabular}{||c|c|c||}
\hline \hline
{name} & {symbol} & {value}\\
    
\hline\hline
FSO link distance	& $ d $ &	$ 2000 $ m  \\

Refractive index structure parameter  & $ C_n^2 $ &	$ 6 \times 10^{-15} $ m${^{-2/3}}$ in weak  turbulence \\

Refractive index structure parameter & $ C_n^2 $ &	$ 2 \times 10^{-14} $ m${^{-2/3}}$ in moderate turbulence \\

Refractive index structure parameter  & $ C_n^2 $ &	$ 5 \times 10^{-14} $ m${^{-2/3}}$ in strong turbulence \\

Optical wavelength	& $ \lambda $ &	$ 1.55~\mu{\rm m} $  \\

Radius of a circular detector aperture & $ a $ &	$ 5 $ cm  \\

Optical beam radius at the waist& $ a_0 $ &	$ 5 $ cm \\

Pointing error (jitter) standard deviation& $\sigma_s $ &$ 5 $ cm \\

Number of relays & $ M $ &	$ 2 $  \\

Order of selected relay	& $ l $ &	$ 2 $   \\
\hline\hline
\end{tabular}

\vspace*{6pt}


\end{table*}

The parameter $ a_d $ is related to the optical beam radius  at the waist, $ a_0 $, and to the radius of curvature, $ F_0 $, as  ${a_d}\!=~\!{a_0}{\left( {({\Theta _o} + {\Lambda _o})(1 + 1.63\chi _R^{12/5}{\Lambda _1})} \right)^{1/2}}$, where $ {\Theta _o} =~1~-~{d \mathord{\left/
 {\vphantom {L {{F_0}}}} \right.
 \kern-\nulldelimiterspace} {{F_0}}}$, $ {\Lambda _o} = {{2d} \mathord{\left/
 {\vphantom {{2d} {(\iota a_0^2)}}} \right.
 \kern-\nulldelimiterspace} {(\iota a_0^2)}}$,  and $ {\Lambda _1} = {{{\Lambda _o}} \mathord{\left/
 {\vphantom {{{\Lambda _o}} {(\Theta _o^2 + \Lambda _o^2)}}} \right.
 \kern-\nulldelimiterspace} {(\Theta _o^2 + \Lambda _o^2)}} $ \cite{PE_MISO}. As it has been mentioned, a standard deviation of pointing errors appears in (\ref{ksi}). By varying this parameter, it is possible to model situation when the alignment between transmitter and receiver is almost perfect. On the other hand, it is also possible to increase standard deviation of pointing errors and describe correctly the situation when tracking is not so precise. 

Based on (\ref{PDF_I}), the electrical SNR is found as  ${\mu _2}=~{{{\eta ^2}P_t^2{\kappa ^2}A_0^2} \mathord{\left/
 {\vphantom {{{\eta ^2}P_t^2{\kappa ^2}A_0^2} {\sigma _{RD}^2}}} \right.
 \kern-\nulldelimiterspace} {\sigma _{RD}^2}} $, with $ \kappa  = {{{\psi ^2}} \mathord{\left/
 {\vphantom {{{\psi ^2}} {({\psi ^2} + 1)}}} \right.
 \kern-\nulldelimiterspace} {({\psi ^2} + 1)}} $.  After some mathematical manipulations and utilizing (\ref{SNR_2}) and (\ref{PDF_I}), the PDF of $ \gamma_{2(l)} $ is derived as \cite{JSAC2}
\begin{equation}
  {f_{{\gamma _{2(l)}}}}({\gamma _2}) = \frac{{{\psi ^2}}}{{2\Gamma (\alpha )\Gamma (\beta ){\gamma _2}}} \MeijerG*{3}{0}{1}{3}{\psi^2+1}{\psi^2 , \, \alpha, \, \beta}{\alpha \beta \kappa \sqrt {\frac{{{\gamma _2}}}{{{\mu _2}}}}}. 
 \label{PDF_SNR2}
\end{equation}

System and channels parameters values, unless otherwise is stated in Numerical results section, are presented in Table I.

\section{System performance analysis}
\label{SecIII}
This section presents the analysis of the system described in Section \ref{SecII} with the aim of deriving analytical expressions for average BER and ergodic capacity. Derivations of average BER and ergodic capacity are based on knowing CDF of the equivalent SNR. This CDF is defined as
\begin{equation}
\begin{split}
{F_{eq}}\left( {{\gamma _{th}}} \right) &= \Pr \left( \frac{{{\gamma _{2(l)}}{\gamma _{1(l)}}}}{{{\gamma _{2(l)}} + \Re}} < \gamma _{th}  \right) \\ 
 &  = \int\limits_0^\infty  {\Pr \left( {{\gamma _{1(l)}} < {\gamma _{th}} + \frac{{{\gamma _{th}}\Re}}{{{\gamma _{2(l)}}}}} \right)} {f_{{\gamma _{2(l)}}}}\left( {{\gamma _{2(l)}}} \right)d{\gamma _{2(l)}} \\ 
&   = \int\limits_0^\infty  {{F_{{\gamma _{1(l)}}}}\left( {{\gamma _{th}} + \frac{{{\gamma _{th}}\Re}}{x}} \right)} {f_{{\gamma _{2(l)}}}}\left( x \right)dx, \\ 
\end{split} 
 \label{Pout_int}
 \end{equation}
where $ \Pr \left(  \cdot  \right) $ denotes the probability.  Substituting (\ref{CDF_SNR_1}) and (\ref{PDF_SNR2}) into (\ref{Pout_int}), after mathematical derivation presented in detail in \cite{chapter}, the final expression for CDF  is derived as \cite[(17.28)]{chapter} 
\begin{equation}
\begin{split}
 {F_{eq}}\left( {{\gamma _{th}}} \right)& =1 - l{M \choose l}\sum\limits_{i = 0}^{l - 1} {l-1 \choose i} \frac{{{{( - 1)}^i}}}{{(M - l + i + 1)}} \hfill {e^{ - \frac{{(M - l + i + 1){\gamma _{th}}}}{{\left( {(M - l + i)(1 - \rho ) + 1} \right){\mu _1}}}}}\\
&  \times \frac{{{2^{\alpha  + \beta  - 3}}{\psi ^2}}}{{\pi \Gamma (\alpha )\Gamma (\beta )}}  \MeijerG*{6}{0}{1}{6}{{\frac{{{\psi ^2} + 2}}{2}} }{\chi_1}{\frac{{{\alpha ^2}{\beta ^2}{\kappa ^2}(M - l + i + 1){\gamma _{th}}\Re}}{{16{\mu _2}\left( {(M - l + i)(1 - \rho ) + 1} \right){\mu _1}}}},
\end{split} 
 \label{Pout}
\end{equation}
where 
\begin{equation}
{\chi _1} = \begin{array}{*{20}{c}}
  {\frac{{{\psi ^2}}}{2},}&{\frac{\alpha }{2},}&{\frac{{\alpha  + 1}}{2},}&{\frac{\beta }{2},}&{\frac{{\beta  + 1}}{2},}&0 .
\end{array}
 \label{chi_1}
\end{equation}

If it is assumed that the pointing errors are small and negligible, it holds that the intensity fluctuations result only from  atmospheric turbulence. In that case, the CDF is derived by taking the limit of (\ref{Pout}) by using
\cite[(07.34.25.0007.01), (07.34.25.0006.01), and (06.05.16.0002.01))]{Wolfarm} and utilizing $ \mathop {\lim }\limits_{\xi  \to \infty } \left( {1 + {2 \mathord{\left/
 {\vphantom {2 {{\xi ^2}}}} \right.
 \kern-\nulldelimiterspace} {{\xi ^2}}}} \right) = 1 $ and $ \mathop {\lim }\limits_{\xi  \to \infty } {\kappa ^2} = \mathop {\lim }\limits_{\xi  \to \infty } {\left( {1 + {1 \mathord{\left/
 {\vphantom {1 {{\xi ^2}}}} \right.
 \kern-\nulldelimiterspace} {{\xi ^2}}}} \right)^2} = 1 $. Obtained expression for the CDF (i.e., outage probability) is reported in \cite[(15)]{JLT}.

\subsection{Average BER}
In the following analysis, the average BER expressions are derived in the case when two modulation formats are applied. More precisely, binary phase-shift keying (BPSK) or differential BPSK (DBPSK) \cite{Alouni} is applied over RF link and SIM-BPSK or SIM-DBPSK \cite{bookMATLAB} is applied over FSO link.  
Following \cite{Ansari-Impact, Ansari-VTCSpring,BER}, the average BER of the system under investigation can be found as 
\begin{equation}
{P_b} = \frac{{{q^p}}}{{2\Gamma (p)}}\int\limits_0^\infty  {{e^{ - q\gamma }}{\gamma ^{p - 1}}{F_{eq}}} \left( \gamma  \right)d\gamma , 
 \label{BER_int}
\end{equation}
where $ {F_{eq}}\left( \gamma  \right) $ is the derived CDF given by (\ref{Pout}), and the parameters $ p $ and $ q $  are $ (p, q) = (0.5, 1) $ for BPSK and SIM-BPSK; $ (p, q) = (1, 1) $ for DBPSK and SIM-DBPSK.

Substituting (\ref{Pout}) into (\ref{BER_int}), the average BER is obtained as 
\begin{equation}
\begin{split}
  {P_b}&= \frac{{{q^p}}}{{2\Gamma (p)}}\int\limits_0^\infty  {{e^{ - q\gamma }}{\gamma ^{p - 1}}\left\{ 1 - l{M \choose l}\sum\limits_{i = 0}^{l - 1} {l-1 \choose i}\frac{{{{( - 1)}^i}}}{{(M - l + i + 1)}}  \right.}  \hfill \\
&   \times {e^{ - \frac{{(M - l + i + 1){\gamma}}}{{\left( {(M - l + i)(1 - \rho ) + 1} \right){\mu _1}}}}} \frac{{{2^{\alpha  + \beta  - 3}}{\psi ^2}}}{{\pi \Gamma (\alpha )\Gamma (\beta )}} \hfill \\
 & \left. { \times\MeijerG*{6}{0}{1}{6}{{\frac{{{\psi ^2} + 2}}{2}} }{\chi_1}{\frac{{{\alpha ^2}{\beta ^2}{\kappa ^2}(M - l + i + 1){\gamma _{th}}\Re}}{{16{\mu _2}\left( {(M - l + i)(1 - \rho ) + 1} \right){\mu _1}}}}} \right\}d\gamma  \hfill \\
 &  = {\Im _1} - {\Im _2} .
\end{split} 
 \label{BER_int1}
\end{equation}

The first integral in (\ref{BER_int1}) is defined and solved using \cite[(3.351.3)]{Grad} as
\begin{equation}
{\Im _1} = \frac{{{q^p}}}{{2\Gamma (p)}}\int\limits_0^\infty  {{e^{ - q\gamma }}{\gamma ^{p - 1}}d\gamma }  = \frac{1}{2}.
 \label{Int_1}
\end{equation}

After transforming the exponential function into Meijer's $ G $-function by utilizing \cite[(01.03.26.0004.01)]{Wolfarm}, the  integral $ \Im _2 $ is expressed as
\begin{equation}
\begin{split}
 {\Im _2} &  = \frac{{{q^p}}}{{2\Gamma (p)}}l{M \choose l}\sum\limits_{i = 0}^{l - 1} {l-1 \choose i} \frac{{{{( - 1)}^i}}}{{(M - l + i + 1)}}  \frac{{{2^{\alpha  + \beta  - 3}}{\psi ^2}}}{{\pi \Gamma (\alpha )\Gamma (\beta )}} \hfill \\
& \times\int\limits_0^\infty  {{\gamma ^{p - 1}} }  \MeijerG*{1}{0}{0}{1}{{{{-}}} }{0}{\left({q + \frac{{(M - l + i + 1)}}{{\left( {(M - l + i)(1 - \rho ) + 1} \right){\mu _1}}}}\right)\gamma}   \hfill\\
& \times\!  { \MeijerG*{6}{0}{1}{6}{{{{{\frac{{{\psi ^2} + 2}}{2}}}}} }{\chi_1}{\frac{{{\alpha ^2}{\beta ^2}{\kappa ^2}(M - l + i + 1)\Re\gamma }}{{16{\mu _2}\left( {(M - l + i)(1 - \rho ) + 1} \right){\mu _1}}}} } d\gamma.
\end{split} 
 \label{Int_2_1}
\end{equation}
After using \cite[(07.34.21.0013.01)]{Wolfarm}, the integral in (\ref{Int_2_1}) is obtained as
\begin{equation}
\begin{split}
 {\Im _2}&  = \frac{{{2^{\alpha  + \beta  - 4}}{\psi ^2}}}{{\pi \Gamma (\alpha )\Gamma (\beta )\Gamma (p)}}l{M \choose l}\sum\limits_{i = 0}^{l - 1} {l-1 \choose i}   \hfill \\
& \! \! \times \frac{{{{( - 1)}^i}}}{{(M - l + i + 1)}}{\left( \!{1\!\! +\!\! \frac{{(M - l + i + 1)}}{{\left( {(M - l + i)(1 - \rho ) + 1} \right)q{\mu _1}}}} \right)^{ - p}} \hfill \\
 &  \times { \MeijerG*{6}{1}{2}{6}{{{{{{1 - p,} \frac{{{\psi ^2} + 2}}{2}}}}} }{\chi_1}{{\frac{{{\alpha ^2}{\beta ^2}{\kappa ^2}\Re}}{{16{\mu _2}\left( {1 + \frac{{q{\mu _1}\left( {(M - l + i)(1 - \rho ) + 1} \right)}}{{(M - l + i + 1)}}} \right)}}}} }\!.
\end{split} 
 \label{Int_2}
\end{equation}

Substituting (\ref{Int_1}) and (\ref{Int_2}) into (\ref{BER_int1}), the final average BER expression is obtained as
\begin{equation}
\begin{split}
&  {P _b} = \frac{1}{2}-\frac{{{2^{\alpha  + \beta  - 4}}{\psi ^2}}}{{\pi \Gamma (\alpha )\Gamma (\beta )\Gamma (p)}}l{M \choose l}\sum\limits_{i = 0}^{l - 1} {l-1 \choose i}   \hfill \\
& \!\times \frac{{{{( - 1)}^i}}}{{(M - l + i + 1)}}{\left( {1 \!\!+ \!\!\frac{{(M - l + i + 1)}}{{\left( {(M - l + i)(1 - \rho ) + 1} \right)q{\mu _1}}}} \!\right)^{\!- p}} \hfill \\
 &  \times { \MeijerG*{6}{1}{2}{6}{{{{{{1 - p,} \frac{{{\psi ^2} + 2}}{2}}}}} }{\chi_1}{{\frac{{{\alpha ^2}{\beta ^2}{\kappa ^2}\Re}}{{16{\mu _2}\left( {1 + \frac{{q{\mu _1}\left( {(M - l + i)(1 - \rho ) + 1} \right)}}{{(M - l + i + 1)}}} \right)}}}} }.
\end{split} 
 \label{BER}
\end{equation}
\\
Under the assumption that the pointing errors are neglected, and the intensity fluctuations are caused only from atmospheric turbulence, the average BER can be obtained by taking the limit of (\ref{BER}), which presents the average BER expression already reported in \cite[(26)]{JLT}.

\subsection{Ergodic capacity}

The assumption is that interleaving is applied at the input of mixed RF-FSO link. This interleaving ensures the FSO channel scintillation remains constant over a frame of symbols and changes for neighboring blocks based on Gamma-Gamma PDF. Similarly, RF channel fading is constant over a frame and changes from one to the next frame based on Rayleigh PDF. In addition, a Gaussian codebook is at the channel input. This codebook is long enough to enable scintillation/fading to be properly described by their PDFs. The ergodic capacity of this composite channel tells us that the maximum information transmission rate is when error probability can be arbitrary low. This is theoretical limit and could be achieved only under previously mentioned conditions. This ergodic capacity should be understood as a benchmark for a given composite RF-FSO channel. Some details related with designing of interleaver depth, which will be sufficient to ensure statistical independence of scintillation/fading from frame to frame, are given in Subsection II.\textit{C}.

For the system under investigation, the ergodic channel capacity, which is determined as $ \hat C\!\!=\! {\rm E}\left[ {{{\log }_2}\left( {1 + e/\left(2\,\pi\right)\,\gamma } \right)} \right] $ in bits/s/Hz \cite{ansari13,R2} (and see references therein), can be derived as 
\begin{equation}
\hat C = B \int\limits_0^\infty  {{{\log }_2}\left( {1 + e/\left(2\,\pi\right)\,\gamma } \right){f_{{\gamma _{eq}}}}\left( \gamma  \right){\kern 1pt} } d\gamma,
 \label{cap_int1}
\end{equation}
where a channel bandwidth is denoted by $ B $, and $ {f_{{\gamma _{eq}}}} \left(  \cdot  \right)$ represents the PDF of overall SNR at the destination. 
Using integration by parts, the ergodic channel capacity in (\ref{cap_int1}) can be presented in terms of the complementary CDF (CCDF) as \cite{cap}
\begin{equation}
\hat C = B\frac{1}{{\ln 2}}\int\limits_0^\infty  {\frac{e}{2\pi}\frac{{F_{\gamma _{eq}} ^c\left( \gamma  \right)}}{{1 + e/\left(2\,\pi\right)\,\gamma }}} d\gamma,
 \label{cap_int2}
\end{equation}
where $ {F_{\gamma_{eq}} ^c\left( \gamma  \right)} $ is the CCDF of overall SNR defined as $ {F_{\gamma_{eq}} ^c\left( \gamma  \right)}=1-{F_{\gamma_{eq}}\left( \gamma  \right)} $ $( {F_{\gamma_{eq}}}$ is the CDF in (\ref{Pout})$ ) $. After substituting (\ref{Pout}) into  (\ref{cap_int2}), and after applying  \cite[(01.02.26.0007.01)]{Wolfarm} 
\begin{equation}
{(1 + e/\left(2\,\pi\right)\,\gamma )^{ - 1}} = \frac{1}{{\Gamma \left( 1 \right)}}\MeijerG*{1}{1}{1}{1}{{{{0}}} }{0}{e/\left(2\,\pi\right)\,\gamma},
 \label{1+x}
\end{equation}
and \cite[(01.03.26.0004.01)]{Wolfarm} 
\begin{equation}
\begin{split} 
{e^{ - \frac{{(M - l + i + 1)\gamma }}{{\left( {(M - l + i)(1 - \rho ) + 1} \right){\mu _1}}}}} \hfill = \MeijerG*{1}{0}{0}{1}{{{{-}}}}{0}{\frac{{(M - l + i + 1)\gamma }}{{\left( {(M - l + i)(1 - \rho ) + 1} \right){\mu _1}}}} ,
\end{split} 
 \label{exp}
\end{equation}
the ergodic capacity in (\ref{cap_int2}) is expressed as
\begin{equation}
\begin{split}
  \hat C &= B\frac{{{2^{\alpha  + \beta  - 3}}{\psi ^2}}}{{\ln 2\pi \Gamma (\alpha )\Gamma (\beta )}}l{M \choose l}\sum\limits_{i = 0}^{l - 1} {l-1 \choose i}\frac{{{{( - 1)}^ie}}}{{(M - l + i + 1)2\pi}}  \hfill \\
 &\times\int\limits_0^\infty\!\! \MeijerG*{1}{1}{1}{1}{{{{0}}}}{0}{\frac{e}{2\,\pi}\,\gamma}\hfill \MeijerG*{1}{0}{0}{1}{{{{-}}}}{0}{\frac{{(M - l + i + 1)\gamma }}{{\left( {(M - l + i)(1 - \rho ) + 1} \right){\mu _1}}}}  \\
&   \times \MeijerG*{6}{0}{1}{6}{{\frac{{{\psi ^2} + 2}}{2}} }{\chi_1}{\frac{{{\alpha ^2}{\beta ^2}{\kappa ^2}(M - l + i + 1){\gamma }\Re}}{{16{\mu _2}\left( {(M - l + i)(1 - \rho ) + 1} \right){\mu _1}}}}d\gamma.
\end{split} 
 \label{cap_int3}
\end{equation}

The integral in (\ref{cap_int3}) can be solved by using \cite[(12)]{gupta}. The final ergodic capacity is obtained in terms of the EGBMGF \cite{BER} as

\begin{equation}\label{cap_ex}
\begin{split}
  \hat C &= B\frac{{{2^{\alpha  + \beta  - 3}}{\psi ^2}}}{{\ln 4\pi^2 \Gamma (\alpha )\Gamma (\beta )}}l{M \choose l}\sum\limits_{i = 0}^{l - 1} {l-1 \choose i} \frac{{{{( - 1)}^ie}((M - l + i)(1 - \rho ) + 1){\mu _1}}}{{{{(M - l + i + 1)}^2}}}\\
   &\times G_{\,1,0:1,1:1,6}^{1,0:1,1:6,0}\left( {\left.\!\!\! {\begin{array}{*{20}{c}}
  1 \\ 
   -  
\end{array}}\!\! \!\right|\left.\!\! {\begin{array}{*{20}{c}}
  0 \\ 
  0 
\end{array}}\!\! \right|\begin{array}{*{20}{c}}
  {\frac{{{\psi ^2} + 2}}{2}} \\ 
  {{\chi _1}} 
\end{array}\left| {{\rm A},{\rm B}} \right.} \right),
\end{split}
\end{equation} 
\\
where 
$
{\rm A} = {\frac{{\left( {(M - l + i)(1 - \rho ) + 1} \right)e{\mu _1}}}{{(M - l + i + 1)2\,\pi}}}
 \label{A}
$ 
and
$  
{\rm B} = {\frac{{{\alpha ^2}{\beta ^2}{\kappa ^2}\Re}}{{16{\mu _2}}}}.
$

\section{Numerical results and simulations}
\label{SecIV}

Based on the derived expressions for the average BER and the ergodic capacity, we obtain numerical results, which are validated via Monte Carlo simulations. The expression  in  (\ref{cap_ex}) is calculated by using the MATHEMATICA® implementation of the EGBMGF given in \cite[Table II]{BER}. Atmospheric turbulence strength is determined by the refractive index structure parameter  for different conditions: $ C_n^2~=~ 6~\times~10^{-15}$~m$ ^{-2/3} $ for weak, $ C_n^2 = 2\times10^{-14}$~m$ ^{-2/3} $ for moderate, and $ C_n^2~=~5\times~10^{-14}$~m$ ^{-2/3} $ for strong turbulence conditions.

\begin{figure}[!t]
\centering
\includegraphics[width=3.5in]{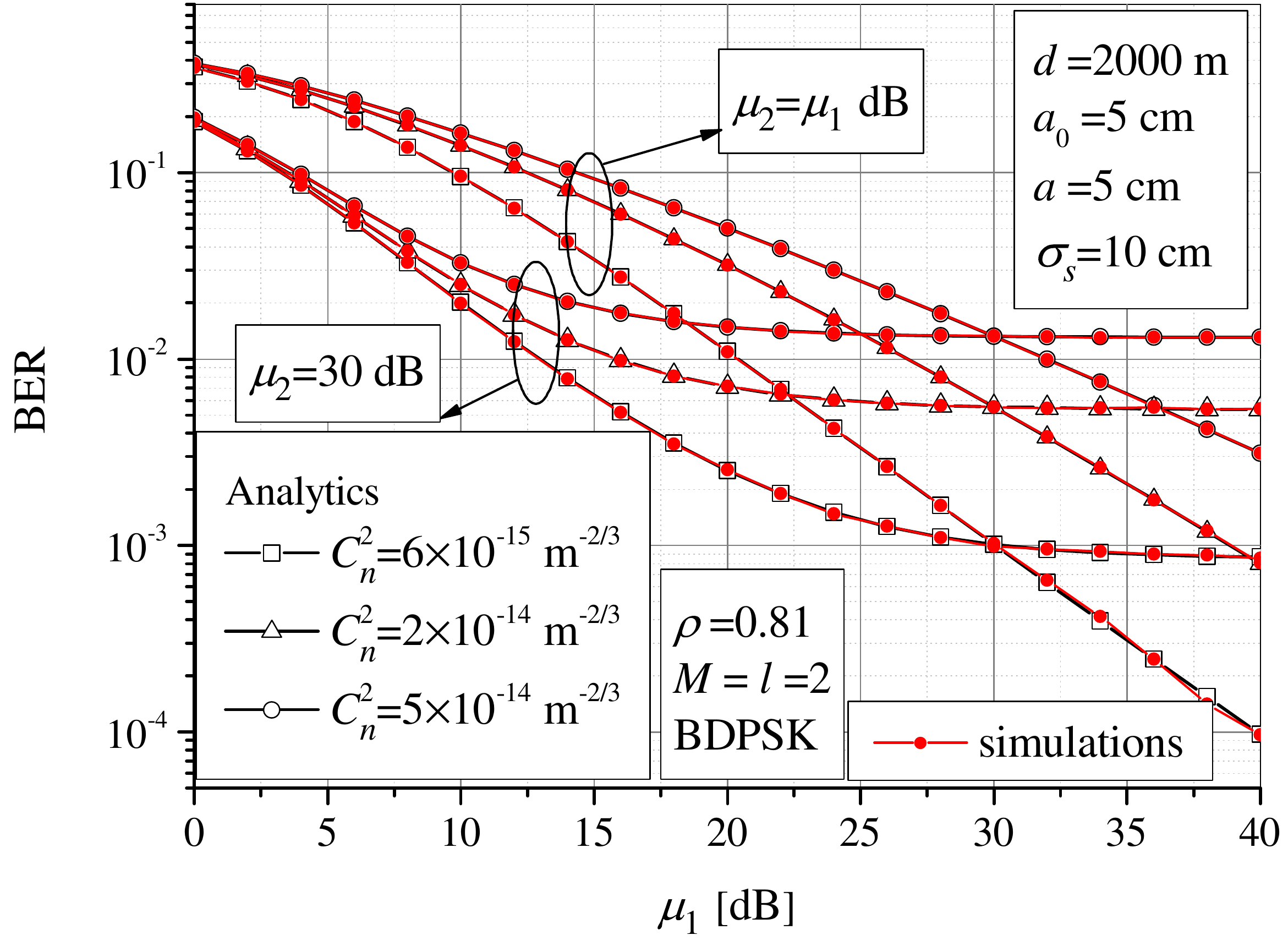}
\caption{Average BER  vs. $\mu_1$ in various atmospheric turbulence conditions}
\label{Fig_2}
\end{figure}

\begin{figure}[!t]
\centering
\includegraphics[width=3.5in]{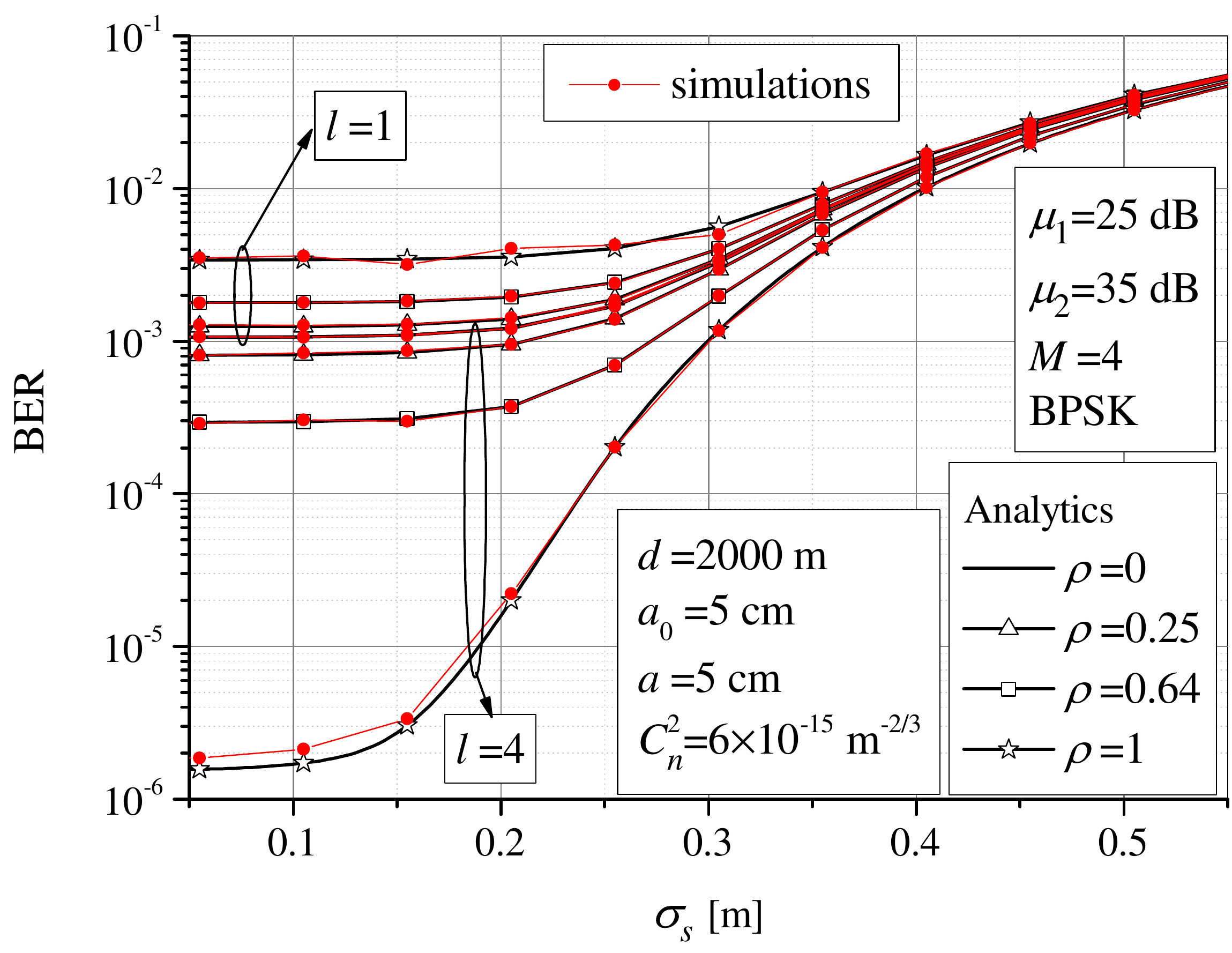}
\caption{Average BER vs. $\sigma _s$ for different values of correlation coefficient}
\label{Fig_3}
\end{figure}

Fig.~\ref{Fig_2} presents the average BER dependence on the average SNR over RF hop in different atmospheric turbulence conditions. Two situations are identified: in the first case $ \mu_2$ has a constant value of 30 dB in the whole range of $ \mu_1$, while in the second case $ \mu_2$ is equal to $ \mu_1$ in the whole range of observation.  As it is expected, system performance is better when the value of $ C_n^2 $ is lower, corresponding to better conditions for the optical signal transmission. When $ \mu_2 $ takes a constant value, the average BER floor occurs, and further increasing the signal power does not improve the system performance. This average BER floor occurs in the range of lower values of $ \mu_1 $ when the FSO hop is under the influence of stronger atmospheric turbulence.


\begin{figure}[!t]
\centering
\includegraphics[width=3.5in]{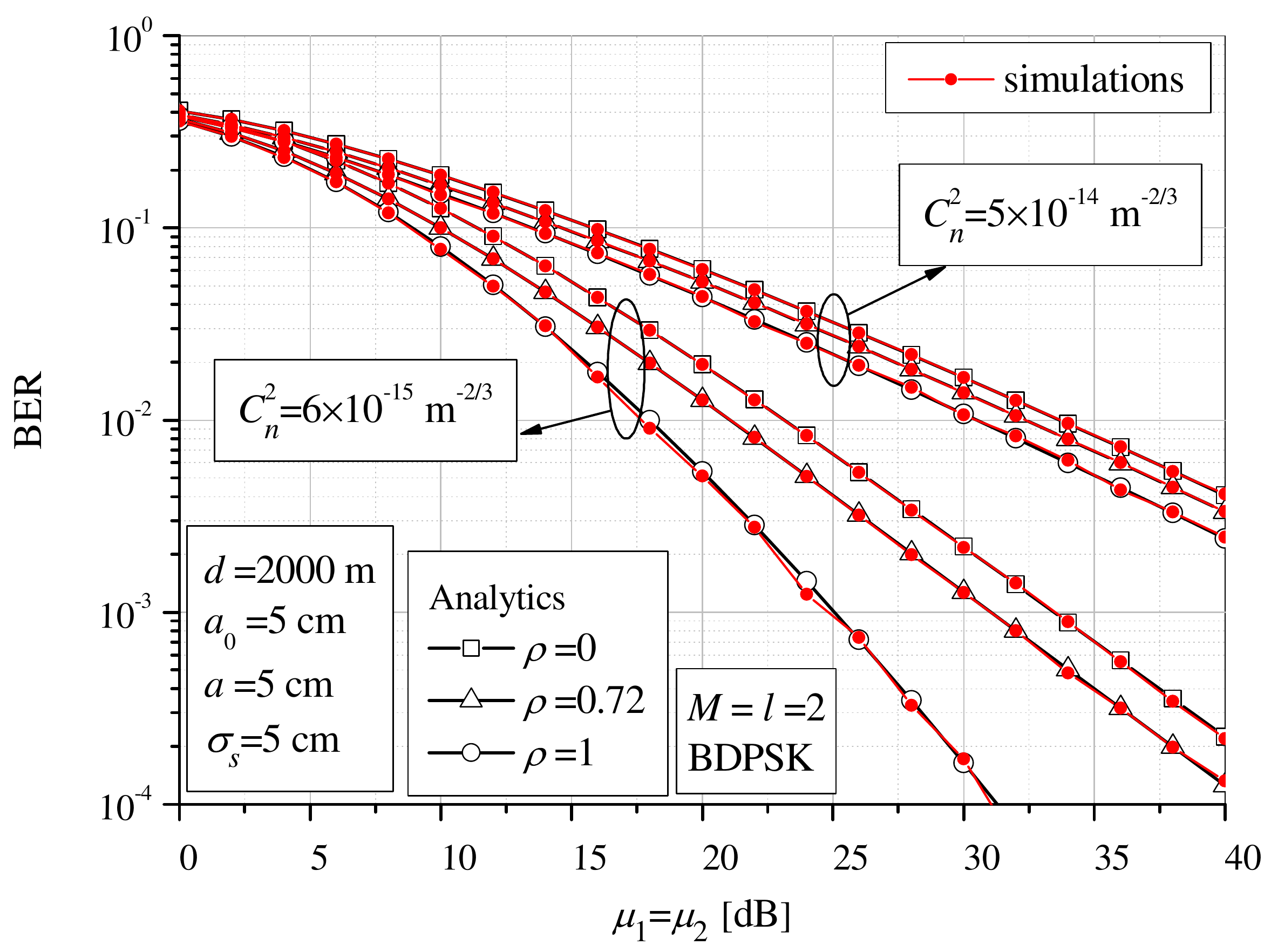}
\caption{Average BER vs. $\mu_1=\mu_2$ in various atmospheric turbulence conditions for different values of correlation coefficient}
\label{Fig_4}
\end{figure}

\begin{figure}[!t]
\centering
\includegraphics[width=3.5in]{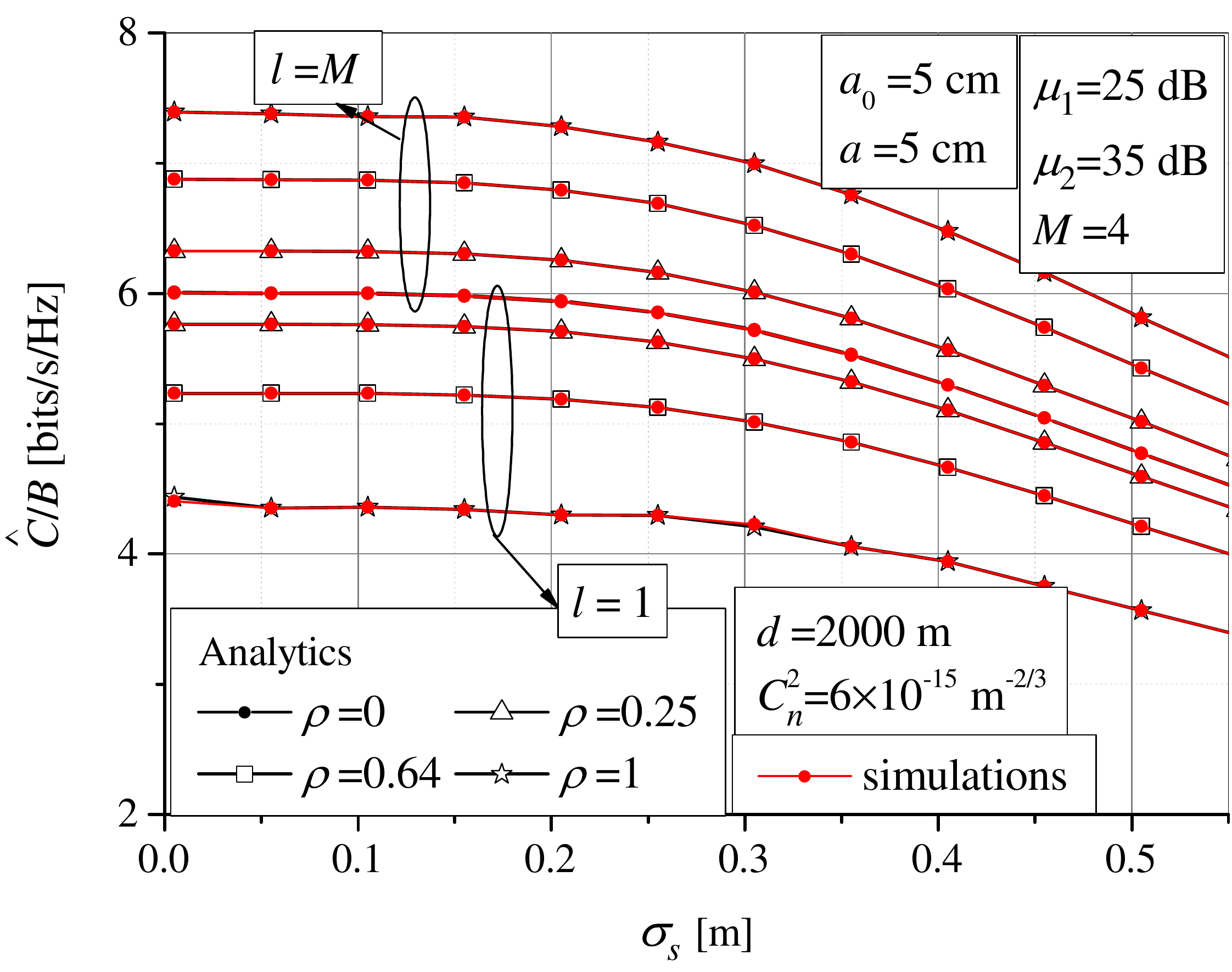}
\caption{Ergodic capacity vs. $\sigma _s$ for different values of correlation coefficient}
\label{Fig_5}
\end{figure}

Fig.~\ref{Fig_3} shows the average BER versus jitter standard deviation when various values of  correlation coefficient are assumed.  The mixed PRS-based RF-FSO system with $ M=4 $ relays is considered. Two scenarios are analyzed: the relay with best estimated CSI can perform further transmission $ (l=M=4) $; and all relays except the one with worst estimated CSI are unavailable $ (l=1) $. 
Greater value of correlation coefficient (meaning the outdated CSI, which is employed for determining of the relay amplification,  and the actual CSI at the time of transmission are more dependent and correlated) leads to better system performance when the best relay transmits the signal. On the other hand, when only the worst one is ready for transmission, greater value of correlation coefficient degrades the system performance. With lowering the correlation coefficient ($ \rho\!\rightarrow\!0 $), outdated and actual CSIs are more independent. In this scenario, it can be decided with high probability that the active relay is not the worst one among all relays, leading to the better system performance. When outdated and actual CSIs are completely uncorrelated, the system performance for the case of $ l=1 $ and $ l=M $ are the same. This occurs since the CSIs are independent and the relay selection has no impact on the system performance.

\begin{figure}[!t]
\centering
\includegraphics[width=3.5in]{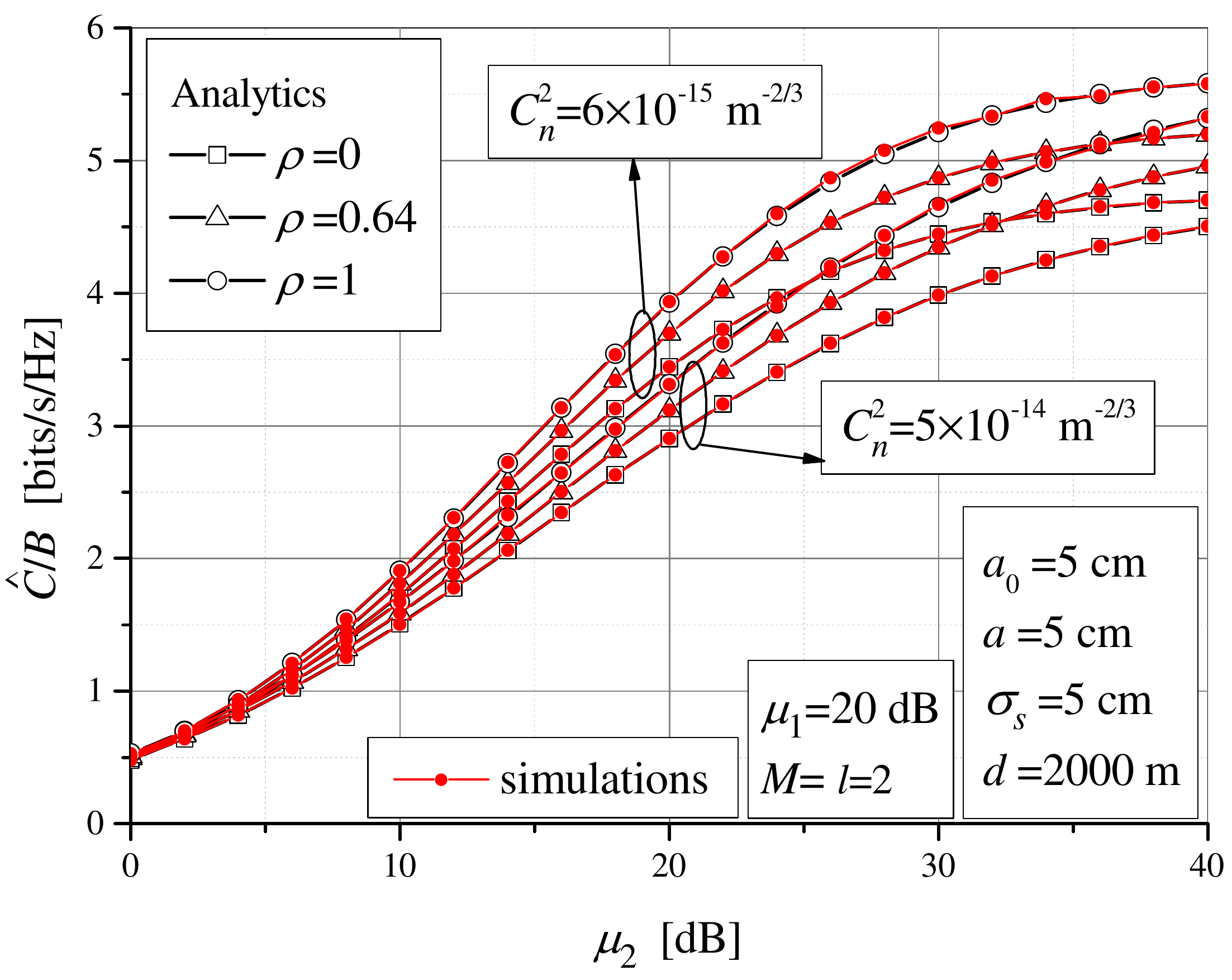}
\caption{Ergodic capacity vs. $\mu_2$ in various atmospheric turbulence conditions for different values of correlation coefficient}
\label{Fig_6}
\end{figure}

Furthermore, Fig.~\ref{Fig_3} shows that the pointing errors  have strong effect on BER performance, especially when the correlation coefficient is greater. Also, the effect of correlation on the average BER is more pronounced when the value of jitter standard deviation is  smaller (corresponding to the weaker pointing errors). In the case of very high values of $ \sigma_s $ $ (\sigma_s>0.4) $, the correlation impact on the RF-FSO system performance is poor and can be neglected. 

Fig.~\ref{Fig_4} presents the average BER in the function of $ \mu_1=\mu_2 $, considering weak and strong atmospheric turbulence and the correlation coefficient $ \rho=0  $, $ \rho=0.72 $, and $ \rho=1 $. It can be observed that greater values of $ \rho $  bring about the improved average BER performance, especially in weak atmospheric turbulence. When second FSO link is affected by convenient conditions (weak atmospheric turbulence), the effect of correlation on the average BER is strong. In the case the transmission of the optical signal  is affected by strong and harmful atmospheric turbulence, the influence of correlation is less significant.

The ergodic capacity versus jitter standard deviation, when various values of correlation coefficient are assumed, is shown in Fig. \ref{Fig_5}. The cases wherein the selected relay is with the best and the worst estimated CSIs, are observed. The same effect as in Fig. \ref{Fig_3} is noticed: the increase of $ \rho $ improves ergodic capacity performance when $ l=M $, while performance degradation is noticed when $ l=1 $. Also, the capacity performance when $ \rho=~0$ is the same for both cases. Contrary to the average BER, it is interesting to note that the pointing errors (determined by $ \sigma_s$) do not play a major role in the ergodic capacity performance. In addition, the intensity of correlation impact on the ergodic capacity is independent of the pointing errors strength.

The ergodic capacity versus the electrical SNR per FSO hop, when the optical signal transmission is performed via channel influenced by various atmospheric turbulence conditions, is presented in Fig.~\ref{Fig_6}. Equivalent to Fig.~\ref{Fig_4}, the capacity performance is better when the FSO link is affected by weak atmospheric turbulence, and when the coefficient correlation is greater. Also, the correlation impact on the ergodic capacity is less dependent on the atmospheric turbulence compared to the average BER performance (see Fig.~\ref{Fig_4}). 

\begin{figure}[!t]
\centering
\includegraphics[width=3.5in]{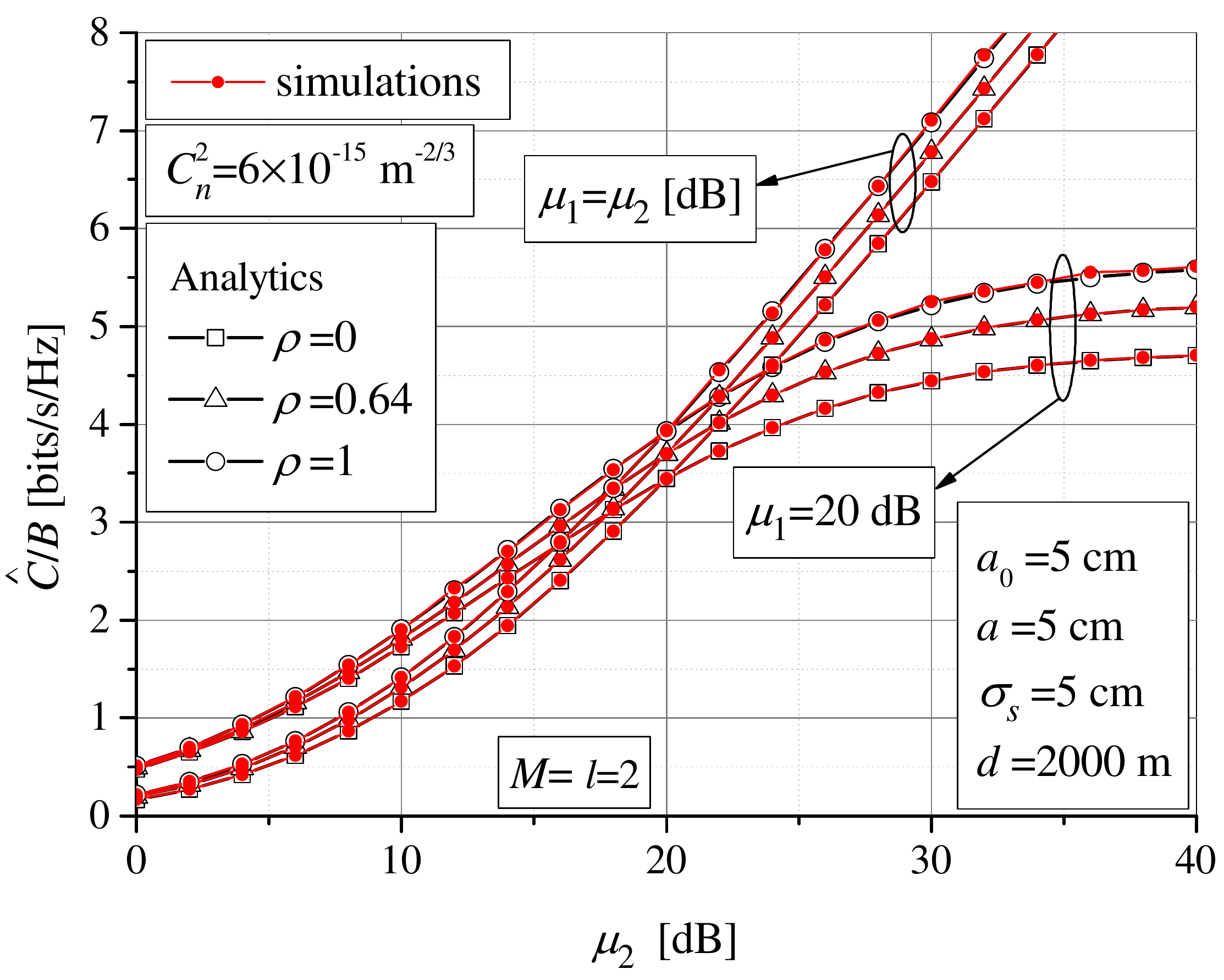}
\caption{Ergodic capacity vs.  $ \mu_2 $ in various atmospheric turbulence conditions for different values of correlation coefficient}
\label{Fig_7}
\end{figure}

The ergodic capacity dependence on the electrical SNR over FSO hop is  depicted in Fig.~\ref{Fig_7}, considering different values of the parameter $ \rho $. The average SNR per RF  link is $ \mu_1=20 $ dB or $ \mu_1=\mu_2 $.  The ergodic capacity performance is improved with greater values of $  \rho $. Similar to Fig.~\ref{Fig_2}, the ergodic capacity floor exists when $ \mu_1 $ is constant, so  the system performance betterment will not be achieved by further increase in the signal power. The capacity floor occurs in the range of lower values of $ \mu_2 $  when $  \rho $ is lower. Contrary, when the average  SNR over RF hop increases simultaneously with the electrical SNR over FSO hop, the ergodic capacity floor does not appear.

\begin{figure}[!t]
\centering
\includegraphics[width=3.5in]{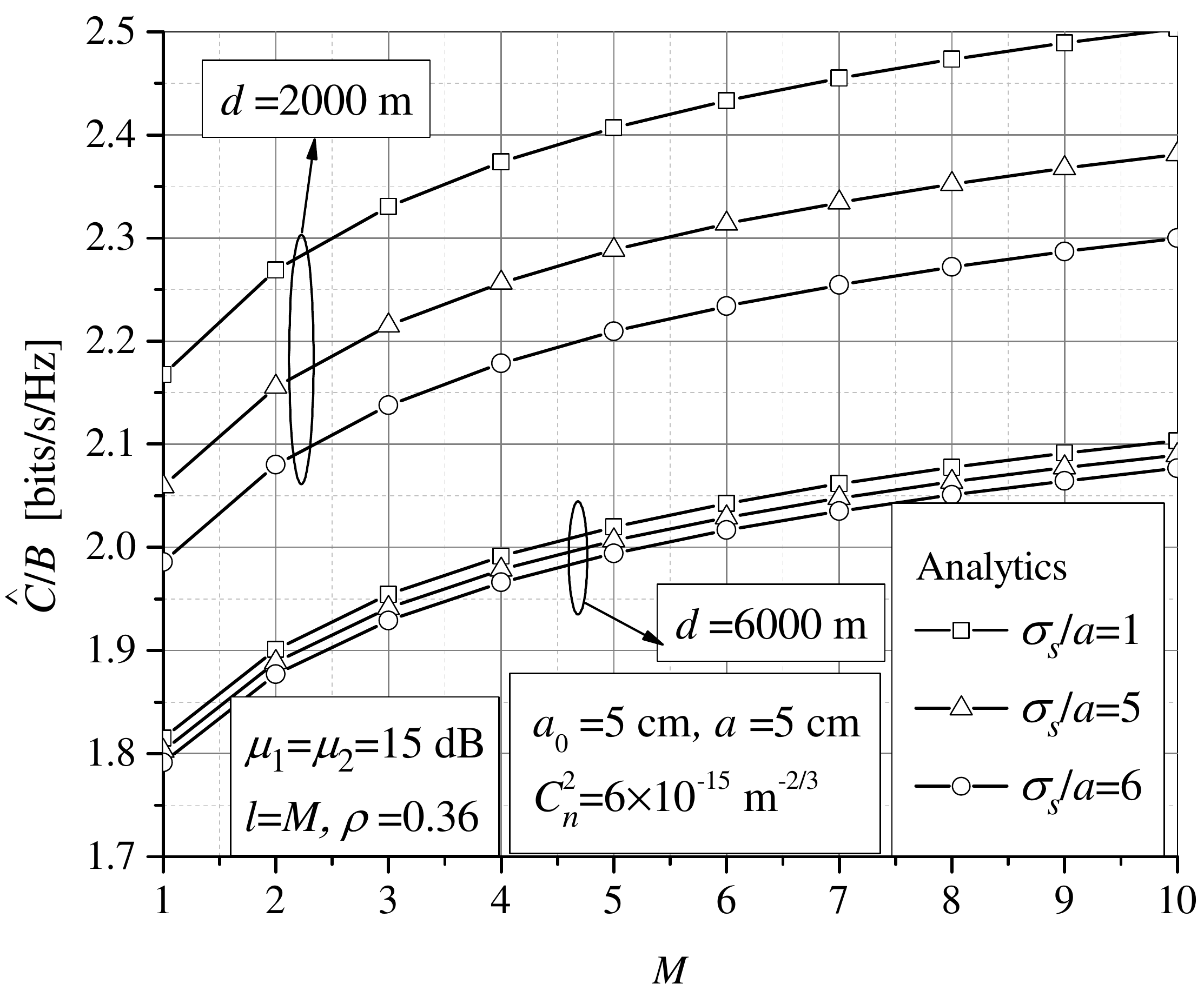}
\caption{Ergodic capacity vs. the number of relays for different values of FSO link length}
\label{Fig_8}
\end{figure}

The ergodic capacity versus the number of relays for  various size of normalized jitter standard deviations is shown in Fig.~\ref{Fig_8}.  It is considered that the range of the FSO hop is $ d=2000 $~m and $ d=6000 $ m. The ergodic capacity performance is better when the FSO link length is shorter, as well as when $ \sigma_s/a $ is lower,  which corresponds to weaker effect of the  misalignment fading. Furthermore, the effect of pointing errors is more dominant when the propagation distance from relay to destination is shorter thereby implying favorable FSO channel conditions.

\begin{figure}[!t]
\centering
\includegraphics[width=3.5in]{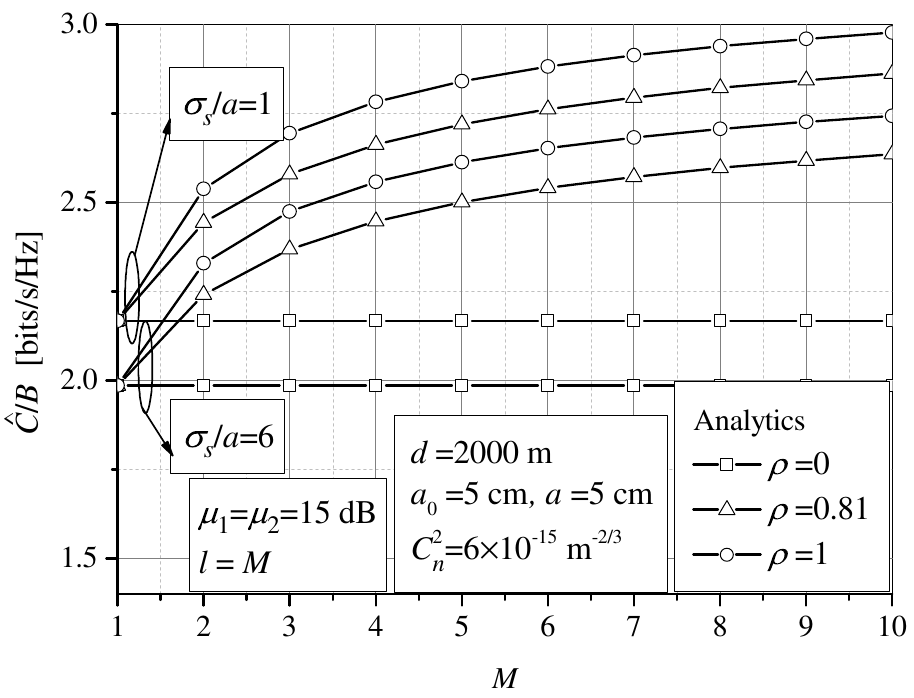}
\caption{Ergodic capacity vs. the number of relays for different values of correlation coefficient}
\label{Fig_9}
\end{figure}

The ergodic capacity dependence on the number of relays for different values of the parameters $ \rho $ and $ \sigma_s/a $ is shown in Fig.~\ref{Fig_9}. When $ \rho=0 $, the outdated and actual CSIs are totally  uncorrelated, and the relay selection has no influence on the ergodic capacity performance. For that reason, the constant value of the capacity is obtained when $ \rho=0 $.  Furthermore, it is observed that the effect of correlation on the ergodic capacity is almost independent on the pointing errors strength. 
Also, the greatest SNR gain is achieved by employing the PRS system with  two  relays compared with the one with  only one relay.

\begin{figure}[!t]
\centering
\includegraphics[width=3.5in]{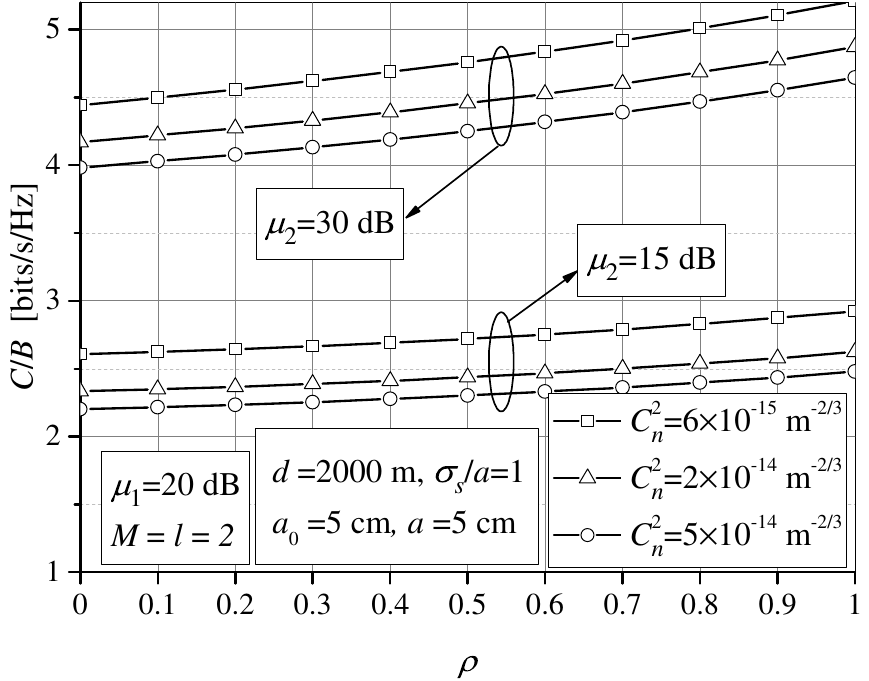}
\caption{Ergodic capacity vs. correlation coefficient}
\label{Fig_10}
\end{figure}

The ergodic capacity versus correlation coefficient is presented in Fig.~\ref{Fig_10}, considering weak, moderate and strong atmospheric turbulence conditions. As it has been concluded, greater values of $ \rho $ lead to improved ergodic capacity performance. In other words, when outdated CSI employed for the relay amplification adjustment and the actual CSI at the time of transmission are more correlated, the value of ergodic capacity is greater. In addition, the slope of capacity curves vs. $\rho$ are the same for all atmospheric turbulence conditions.

\section{Concluding remarks}
\label{SecV}


We have analyzed the average BER and the ergodic capacity dependence on atmospheric turbulence, pointing errors strength, correlation coefficient, electrical SNR per FSO hop, average SNR per RF hop, and different PRS structures. It has been concluded that the temporal correlation coefficient is an important parameter influencing the system performance. 
Greater values of the correlation coefficient (i.e., meaning that the outdated CSI and actual CSI of the source-relay channel at the time of signal transmission are more correlated) lead to improvement of the average BER (ergodic capaciy) performance in the case when the relay with best estimated CSI is available. Contrary, average BER (ergodic capacity) performance becomes worse with increasing correlation coefficient in the case when all relays except the one with the worst estimated CSI are unavailable. When the correlation coefficient is equal to zero, the average BER (ergodic capacity) performance is the same independently if the best or the worst relay is selected.

Furthermore, the impact of correlation on the average BER is more pronounced in the case when the FSO signal experiences friendly environment with favorable conditions (weak pointing errors and weak atmospheric turbulence). On the other hand, the slope of the ergodic capacity curve vs. correlation coefficient is approximately the same in all turbulence conditions of FSO link.
In addition, the following conclusion follows: the larger the value of correlation coefficient, the stronger is the effect of number of relays on the ergodic capacity.

\section*{Acknowledgment }
The work of M. I. Petkovic has received funding from the European Union Horizon 2020 research and innovation programme under the Marie Skodowska-Curie grant agreement No 734331.  This work was partially supported by the Ministry of Education, Science and Technology Development of the Republic of Serbia under Grants TR-32035, TR-32028 and III-44006, as well as COST Action CA16220.


\end{document}